\documentclass[final]{aipproc}

\layoutstyle{6x9}
\def\prl{Phys. Rev. Lett.}

\def\ptrsl{Phil. Trans. Roy. Soc. London}

\def\apj{ApJ}
\def\apjl{ApJL}
\def\pre{Phys. Rev. E}

\begin{document}

\title{Scaling Laws and Intermittency in Highly Compressible Turbulence}



\author{Alexei G. Kritsuk$^{\dagger}$, Paolo Padoan, Rick Wagner, \\and Michael L. Norman}{
address={Physics Department and Center for Astrophysics and Space Sciences,\\
University of California, San Diego,
9500 Gilman Drive, La Jolla, CA 92093-0424, USA}
,altaddress={$^{\dagger}$also Sobolev Astronomical Institute, St. Petersburg State University, St. Petersburg, Russia}
}

\begin{abstract}
We use large-scale three-dimensional simulations of supersonic
Euler turbulence to study the physics of a highly 
compressible cascade. Our numerical experiments describe non-magnetized 
driven turbulent flows with an isothermal equation of state and an rms Mach 
number of~6. We find that the inertial range velocity scaling deviates
strongly from the incompressible Kolmogorov laws. We propose an
extension of Kolmogorov's K41 phenomenology that takes into account
compressibility by mixing the velocity and density statistics and preserves 
the K41 scaling of the 
density-weighted velocity $v\equiv\rho^{1/3}u$. We show that low-order 
statistics of $v$ are invariant with respect to changes in the Mach 
number. For instance, at Mach 6 the slope of the power spectrum of $v$ 
is~$-1.69$ and the third-order structure function of $v$ scales 
linearly with separation.
We directly measure the mass dimension of the ``fractal'' density
distribution in the inertial subrange, $D_m\approx 2.4$, which is similar 
to the observed fractal dimension of molecular clouds and agrees well 
with the cascade phenomenology.
\end{abstract}

\maketitle

\section{Introduction}

In the late 1930's, Kolmogorov clearly realized that chances 
to develop a closed purely mathematical theory of turbulence are 
extremely low \cite{kolmogorov85}.\footnote{``An understanding 
of solutions to the [incompressible] 
Navier-Stokes equations'' yet remains one of the six unsolved grand 
challenge problems nominated by the Clay Mathematics Institute in 
2000 for a \$1M {\em Millennium Prize} [http://www.claymath.org/millennium/].}
Therefore, the basic approach in \cite{kolmogorov41a,kolmogorov41c} 
(usually referred to as the K41 theory) was to rely on physical 
intuition and formulate two general statistical hypotheses which
describe the universal equilibrium regime of small-scale fluctuations
in arbitrary turbulent flow at high Reynolds number. Following the 
Landau (1944) remark on the lack of universality in turbulent flows 
\cite{landau.87}, and with information extracted from
new experimental data, the original similarity hypotheses of K41 
were then revisited and refined to account for 
intermittency effects \cite{kolmogorov62,she.94,dubrulle94}. While 
the K41 phenomenology became the cornerstone for all subsequent 
developments in incompressible turbulence research 
\cite[e.g.,][]{frisch95}, there was no similar result established 
for compressible flows yet \cite{lele94,friedrich07}. Historically,
compressible turbulence research, preoccupied with a variety of 
specific engineering applications, was generally lagging behind 
the incompressible developments.\footnote{A reasonable measure of 
the delay is 60+ years passed between the appearance of incompressible 
Reynolds averaging \cite{reynolds1895} and mass-weighted Favre 
averaging for fluid flows with variable density \cite{favre58},
although see \cite{lumley.01} for references to a few earlier 
papers that dealt with density-weighted averaging.}
The two major reasons for this time lag were an additional complexity of
analytical treatment of compressible flows and a shortage in
experimental data for super- and hypersonic turbulence. In this 
respect, although limited to relatively low Reynolds numbers, 
direct numerical simulations (DNS) of turbulence (pioneered by 
Orszag and Patterson \cite{orszag.72}) have occupied 
the niche of experiments at least for the most simple flows. One
particularly important advantage of DNS is an easy access to 
variables that are otherwise difficult to measure in the 
laboratory or treat analytically.

A traditionally straightforward approach to data analysis from DNS 
of compressible turbulence includes computation of the ``standard'' 
statistics of velocity fluctuations. In addition, the diagnostics 
for density fluctuations are also computed and discussed as the 
direct measures of compressibility. Quite naturally, both density 
and velocity statistics demonstrate strong dependence on the Mach 
number ${\cal M}$ in supersonic (${\cal M}\in [1,3]$) and 
hypersonic (${\cal M}>3$) regimes, while the variations in turbulent 
diagnostics at sub- or transonic Mach numbers are rather small. 
For instance, at ${\cal M}\approx 1$ the velocity power spectrum closely 
follows the K41 scaling and the third order velocity structure 
functions scale roughly linearly with separation \cite{porter..02}.
The density power spectrum in weakly compressible isothermal
flows scales as $\sim k^{-7/3}$ \cite{bayly..92}, at ${\cal M}\approx 1$
it scales as $\sim k^{-1.7}$ \cite{kritsuk...07}, and at ${\cal M}\approx 6$
the slope is $-1.07$ \cite{kritsuk...07}. 

\begin{figure}
  \includegraphics[height=.22\textheight]{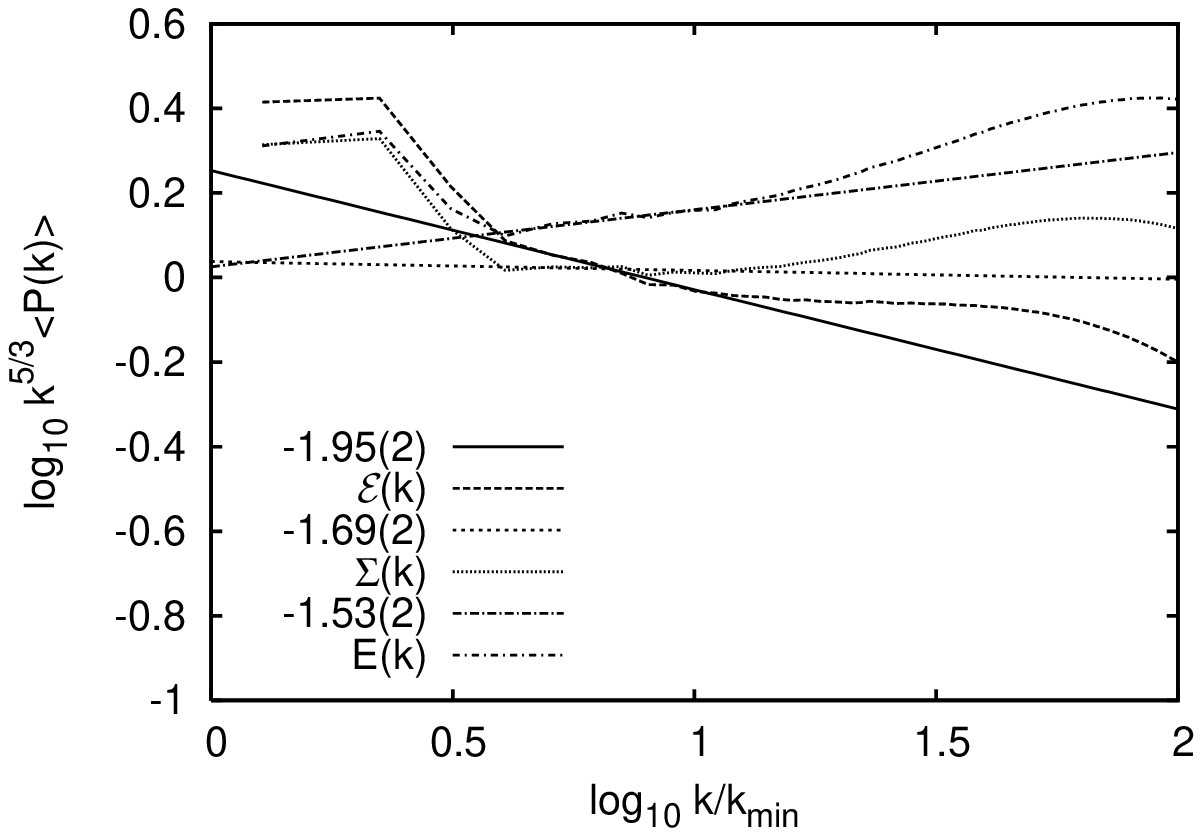}
  \hfill
  \includegraphics[height=.22\textheight]{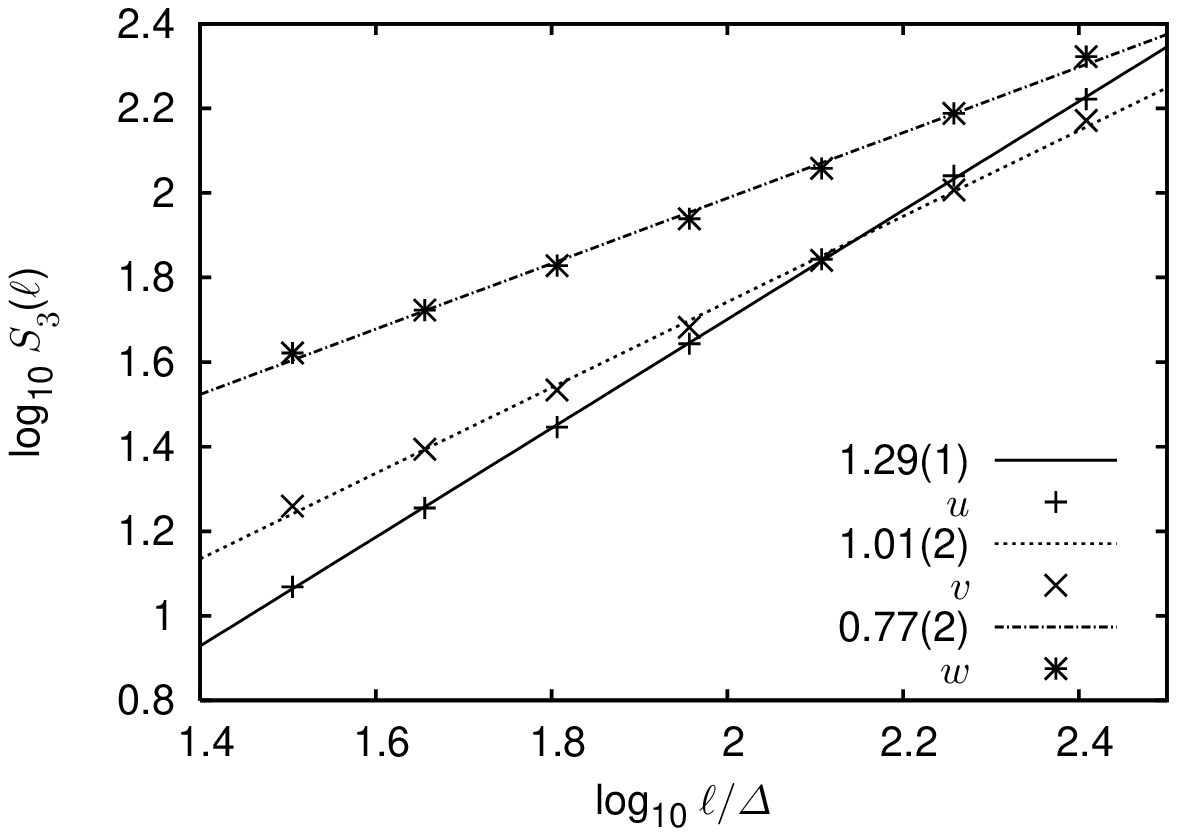}
  \caption{Time average compensated power spectra ({\em left}) and 
third-order transverse structure functions ({\em right}) for velocity 
$u$ and mass-weighted velocities $v\equiv\rho^{1/3}u$ and 
$w\equiv\rho^{1/2}u$. The statistics of $v$ clearly demonstrate 
a K41-like scaling. Notice strong bottleneck contamination in the
spectra at high wavenumbers.}
\label{k41laws}
\end{figure}

Based on the data from numerical experiments, it is well established that: 
(i) the velocity power spectra tend to get steeper as the Mach number increases, 
reaching the Burgers slope of $-2$ asymptotically \cite[and references 
therein]{biskamp03}; (ii) the density power 
spectra instead get shallower at high Mach numbers, approaching a slope
of $-1$ or even shallower \cite{kritsuk..06}; (iii) the density PDF in 
isothermal turbulent flows is well 
represented by a lognormal distribution \cite[and references therein]{biskamp03}; 
(iv) the dimensionality of the most singular velocity structures increases
from $D_{s,u} \sim 1$ in a subsonic regime to $D_{s,u} \sim 2$ 
in highly supersonic \cite{padoan...04}; (v) the mass dimension of the 
turbulent structures decreases from $D_m=3$ in weakly compressible flows 
to $D_m\sim2.5$ in highly compressible \cite{kritsuk...07}.

How can we combine these seemingly disconnected pieces of
information into a coherent physical picture to 
improve our understanding of compressible turbulence? One way to do this
is to consider a phenomenological concept of a {\em lossy} compressible 
turbulent cascade that would asymptotically match the incompressible 
Kolmogorov-Richardson energy cascade \cite{kolmogorov41a,richardson22}
in the limit of very low Mach numbers. Since incompressible turbulence 
represents a degenerate case where the density is uncorrelated with 
the velocity, the phenomenology of the compressible cascade must
include this correlation. This essentially means that instead of 
velocity $u$, which is a single key ingredient of the K41 laws, one 
needs to consider a set of mixed variables, $\rho^{1/\eta}u$, 
where $\rho$ is the density and $\eta$ can take values $1$, $2$, 
or $3$, depending on the statistical measure of interest \cite{kritsuk...07}. 
For instance, if one is studying the scale-by-scale kinetic 
energy budget in a compressible turbulent flow, a mixed variable 
power spectrum with $\eta=2$ would be an appropriate choice. To
deal with the kinetic energy flux through the hierarchy of
scales within the inertial range, the key mixed variable would 
be the one with $\eta=3$.

How will these mixed statistics scale in the inertial range of
highly compressible turbulent flows? Will their scaling depend on
the Mach number? Can the K41 phenomenology be extended to cover
hypersonic turbulent flows? These and other related questions are
in detail discussed in \cite{kritsuk...07} based on Euler
simulations of driven isotropic supersonic turbulence with the Piecewise
Parabolic Method \cite{colella.84} and with resolution up to $2048^3$ grid points. 
In this paper we present the highlights of the compressible 
cascade phenomenology verified in \cite{kritsuk...07}.

\section{Scaling, Structures, and Intermittency} 

Nonlinear interactions transfer kinetic energy supplied to the system 
at large scales through the inertial range with little dissipation. Let 
us assume that the mean {\em volume} energy transfer rate in a 
compressible fluid, $\rho u^2 u/\ell$, is constant in a statistical 
steady state \cite[e.g.,][]{lighthill55}. If this is true, then
\begin{equation}
v^p\equiv(\rho^{1/3}u)^p\sim \ell^{\;p/3}
\label{mxd}
\end{equation}
for an arbitrary power $p$ and, with the standard assumption of 
self-similarity of the cascade, the structure functions (SFs) 
of mixed variable $v$ for compressible flows should scale in the 
inertial range as
\begin{equation}
{\cal S}_p(\ell)\equiv\left<\left|v(r+\ell) - v(r)\right|^p\right>\sim \ell^{\;p/3}.
\label{sf41}
\end{equation}
%

In the limit of weak compressibility, the scaling laws 
(\ref{sf41}) will reduce to the K41 results for the velocity 
structure functions. The scaling laws 
${\cal S}_p(\ell)\sim \ell^{\zeta_p}$, where $\zeta_p=p/3$ 
are not necessarily exact. As the incompressible K41 
scaling, they are subject to ``intermittency corrections'', e.g.
$\zeta_p=p/3+\tau_{p/3}$ \cite{kolmogorov62}. The only exception is, 
perhaps, the third order relation for the longitudinal velocity 
SFs, which is exact in the incompressible case and is known as 
the {\em four-fifth law} \cite{kolmogorov41c}. 
Our focus here is mostly on the low order
statistics ($p\leq3$) for which the corrections are small. Since
the power spectrum slope is related to the exponent of the second
order structure function, the K41 slope of $5/3$ is expected to
hold for $v\equiv\rho^{1/3}u$ in the compressible case.

\begin{figure}
  \includegraphics[height=.22\textheight]{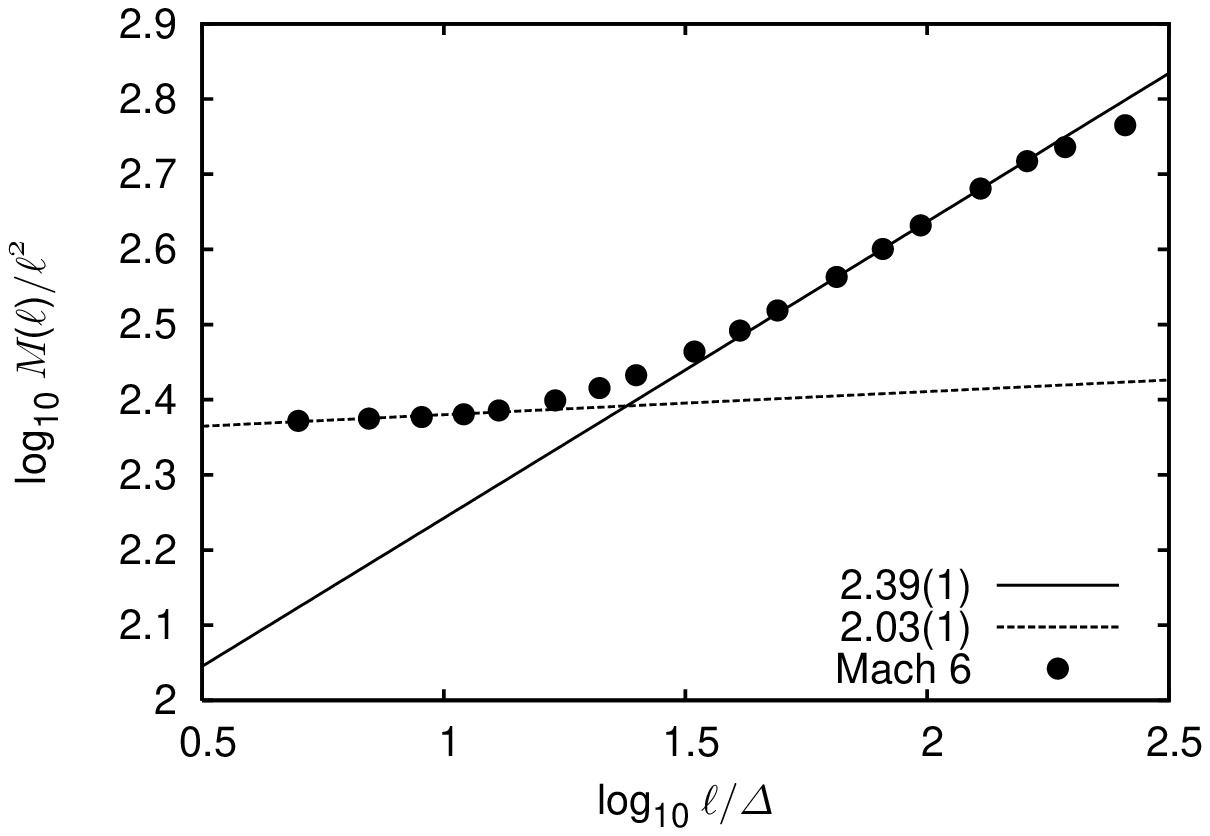}
  \hfill
  \includegraphics[height=.22\textheight]{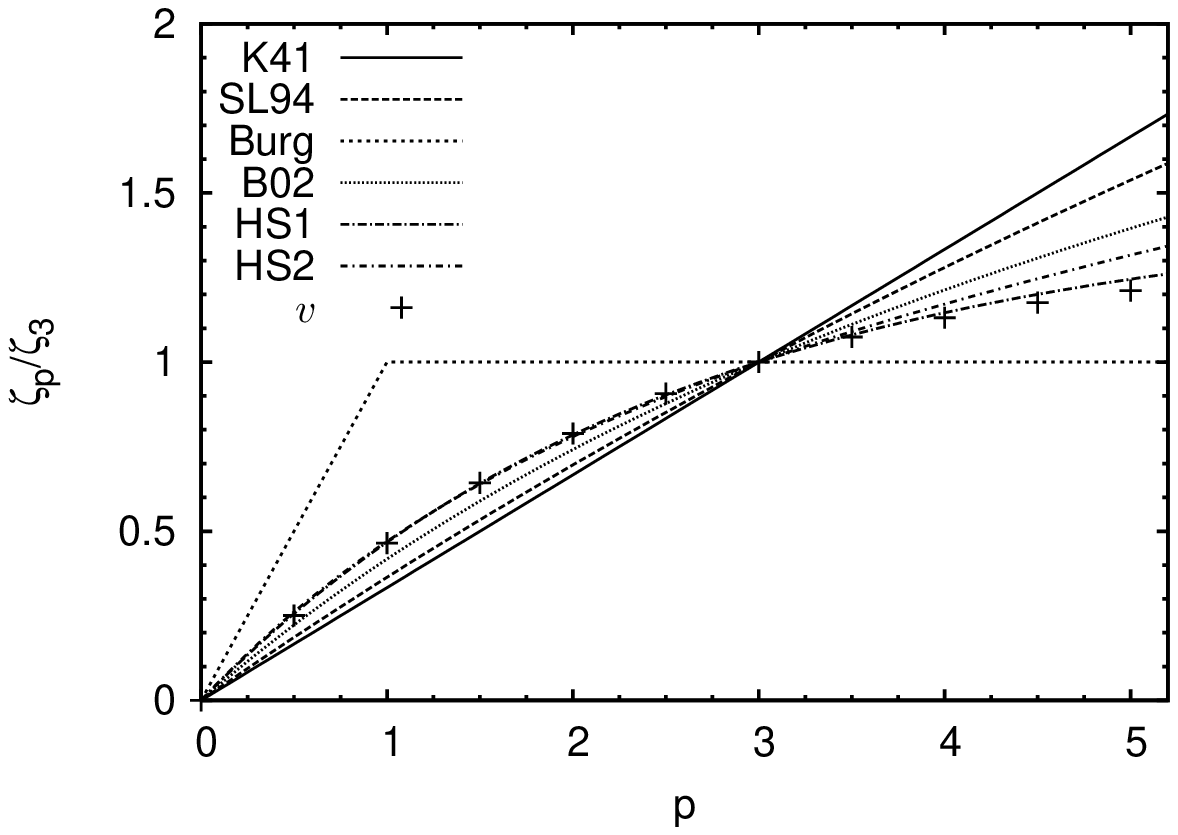}
  \caption{Gas mass $M(\ell)$ as a function of the box size 
$\ell$ ({\em left}). The mass dimension $D_m$ is defined as the 
log-log slope of $M(\ell)$, see eq. (4). Relative exponents for structure
functions of the transverse modified velocities $v$ versus
order $p$ and two hierarchical structure models with different parameters
\cite[HS1 \& HS2,][]{she.94} that fit the data for $p\in[0,\,3]$ ({\em right}). Also shown 
are model predictions for the Kolmogorov-Richardson cascade 
\cite[K41,][]{kolmogorov41a,kolmogorov41c}, for intermittent 
incompressible turbulence \cite[SL94,][]{she.94}, for ``burgulence''
\cite[Burg,][]{bec.07}, and for the velocity fluctuations in 
supersonic turbulence \cite[B02,][]{boldyrev02}.}
\label{mass}
\end{figure}

Figure~\ref{k41laws} shows the power spectra of $u$, $v$, and 
$w\equiv\rho^{1/2}u$ and the corresponding third-order transverse 
structure functions based on the simulations at Mach 6 
\cite{kritsuk...06,kritsuk...07}. The power spectrum 
$\Sigma(k)$ and the structure function of $v$ clearly follow the
K41 scaling: $\Sigma\sim k^{-1.69}$ and ${\cal S}_3\sim\ell^{1.01}$ \cite{kritsuk...07}, 
while the velocity power spectrum ${\cal E}(k)$ and structure function 
have substantially steeper-then-K41 slopes: $-1.95$ and 1.29 \cite{kritsuk...06}. 
At the same time, the kinetic energy spectrum $E\sim k^{-1.53}$ is 
shallow and both solenoidal and dilatational components of $w$ have
the same slope implying a single compressible energy cascade 
with strong interaction between the two components \cite{kritsuk...07}.
These results based on the high dynamic range simulations lend strong 
support to the scaling relations described by eq.~(\ref{sf41}) 
and to the conjecture from which they were inferred. Previous
simulations at lower resolution did not allow to measure the
absolute exponents reliably due to insufficient dynamic range
and due to the bottleneck contamination \cite{zakharov..92}. 

\begin{figure}
  \includegraphics[height=.45\textheight]{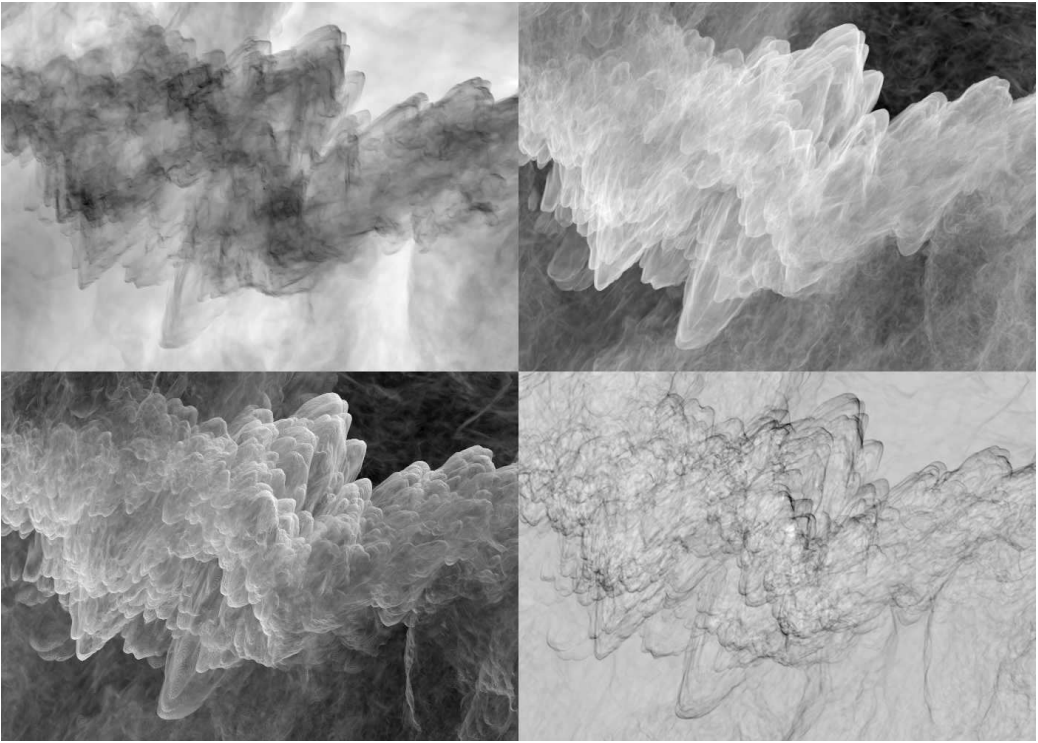}
  \caption{Coherent structures in Mach 6 turbulence at resolution of $1024^3$.
Projections along the minor axis of a subvolume of $700\times500\times250$
zones for the density ({\em upper left}),
the enstrophy ({\em upper right}),
the dissipation rate ({\em lower left}), and the dilatation ({\em lower right}). 
The logarithmic grey-scale ramp shows the lower values
as dark in all cases except for the density. The inertial subrange structures
correspond to scales between $40$ and $250$ zones and represent a fractal
with $D_m\approx2.4$. The dominant structures in the dissipation range
($\ell < 30\Delta$) are shocks with $D_m=2$. [Reprinted from 
\cite{kritsuk...07}.]}
\label{struc}
\end{figure}

In 1951, von Weisz\"aker \cite{vonweizs51} introduced a 
phenomenological model for scale-invariant
hierarchy of density fluctuations in compressible 
turbulence described by a simple equation that relates the
mass density at two successive levels to the corresponding 
scales through a universal measure of the degree of 
compression, $\alpha$,
\begin{equation}
\rho_n/\rho_{n-1}=\left(\ell_n/\ell_{n-1}\right)^{-3\alpha}.
\label{vws}
\end{equation}
The geometric factor $\alpha$ takes the
value of $1$ in a special case of isotropic compression 
in three dimensions, $1/3$ for a perfect one-dimensional 
compression, and zero in the incompressible limit.
From equations (\ref{mxd}) and (\ref{vws}), assuming 
mass conservation, Fleck \cite{fleck96} derived a set of 
scaling relations for the velocity, specific kinetic energy, 
density, and mass:
\begin{equation}
u\sim \ell^{1/3+\alpha},\;\;
{\cal E}(k)\sim k^{\,-5/3-2\alpha},\;\;
\rho\sim \ell^{\;-3\alpha},\;\;
M(\ell)\sim \ell^{D_m}\sim \ell^{\;3-3\alpha},
\label{f96}
\end{equation}
where all the exponents depend on the compression measure 
$\alpha$ which is in turn a function of the rms Mach number 
of the turbulent flow. We can now use the data from numerical 
experiments to verify the scaling relations (\ref{f96}). 
Since the first-order velocity structure function scales as 
$\ell^{0.54}$ \cite{kritsuk...07}, 
we can estimate $\alpha$ for the Mach 6 flow, $\alpha\approx0.21$.
Using the last relation in (\ref{f96}), we can calculate the mass 
dimension for the density distribution, $D_m\approx2.38$. 
It is indeed consistent with our direct measurement of the 
mass dimension for the same range of scales, 
$D_m\approx2.39$, see Fig.~\ref{mass}. 

In strongly compressible turbulence at Mach 6, the density contrast 
between supersonically moving blobs and their more diffuse environment
can be as high as $10^6$. The most common structural elements in
such highly fragmented flows are nested bow-shocks \cite{kritsuk..06}.
Figure~\ref{struc} shows an extreme example of structures
formed by a collision of counter-propagating supersonic flows. On 
small scales within the dissipation range, these structures are 
characterized by $D_m=2$, while within the inertial range 
$D_m\approx2.4$ (Fig.~2, {\em left}). The hierarchical structure (HS)
model
\begin{equation}
\zeta_p/\zeta_3=\gamma p + C(1-\beta^p)
\end{equation}
\cite{she.94} provides
good fits to the data for the mass-weighted velocity $v$ (see Fig.~2, {\em right}). 
Here the codimension of the support of the most singular 
dissipative structures 
\begin{equation}
C\equiv3-D_{s,v}=(1-3\gamma)/(1-\beta^3).
\end{equation}
If the fit is
limited to $p\in[0,\,3]$, two sets of model parameters $\beta$ and $\gamma$ 
are formally acceptable (models HS1 and HS2 in Fig.~2).
The best-fit parameters of the HS1 model: $\beta_1^3=1/3$ (a measure of 
intermittency),
$\gamma_1=0$ (a measure of singularity of structures), and $C_1=1.5$ correspond 
to a hybrid between the B02 model for the velocity fluctuations 
($\beta_{B02}^3=1/3$, $\gamma_{B02}=1/9$) \cite{boldyrev02} and the Burgers' 
model ($\beta_{Burg}=0$, $\gamma_{Burg}=0$) \cite{bec.07}. The HS2 model
($\beta_2^3=1/6$, $\gamma\,_2=1/9$, and $C_2=0.8$) provides a fit of roughly the same 
quality for $p\in[0,\,3]$, but overestimates the scaling exponents $\zeta_p$ at $p>4$.
Since the level of uncertainty in the high order statistics remains 
high even at a resolution of $1024^3$ grid points, larger dynamic 
range simulations are needed to distinguish between the two options.

If the HS1 option is confirmed, then Mach~6 turbulence  
is more intermittent than incompressible turbulence
($\beta_1^3<\beta_{SL94}^3=2/3$) and has the same degree of singularity 
of structures as burgulence. The singular dissipative structures with 
fractal dimension $D_{s,v}=1.5$ can be conceived as perforated sheets
reminiscent of the Sierpinski sieve. 
If the HS2 option is justified, then turbulence at Mach 6 is even more
intermittent, but has the same degree of singularity of structures
as incompressible turbulence ($\gamma_2=\gamma_{SL94}$). In this case the fractal
dimension of the most singular structures, $D_{s,v}=2.2$, is slightly higher
than in the B02 model \cite{boldyrev02}. Formally, it is also possible that both
types of structures are present in highly compressible turbulence, implying 
multiple modulation defects and a compound nature of Poisson statistic 
\cite[cf.][]{she.95}. In this case, a linear combination of HS1 and HS2
models would describe the high order exponents best.

\section{Conclusion}

Using large-scale Euler simulations of supersonic turbulence at Mach 6 we 
have demonstrated
that there exists an analogue of the K41 scaling laws valid for both weakly and
highly compressible flows. The mass-weighted velocity $v\equiv\rho^{1/3}u$
-- the primary variable governing the energy transfer through the cascade -- 
should replace the velocity $u$ in intermittency models for compressible flows
at high Mach numbers.

\begin{theacknowledgments}
This research was partially supported by a NASA ATP grant NNG056601G,
by NSF grants AST-0507768 and AST-0607675, and by NRAC allocations 
MCA098020S and MCA07S014. We utilized computing resources provided by 
the San Diego Supercomputer Center and by the National Center for 
Supercomputer Applications.
\end{theacknowledgments}

\end{document}